\documentclass[showpacs,aps,pra,twocolumn]{revtex4}
\usepackage{graphicx,epsfig}
\usepackage{times}
\usepackage{amsmath,amssymb}

\newlength\figurewidth
\newcommand{\fig}{Fig.}

\newcommand{\sect}{Sec.}

\newcommand{\rem}[1]{}

\newcommand{\imag}[1]{\text{Im}(#1)}

\newcommand{\real}[1]{\text{Re}(#1)}

\newcommand{\realc}[1]{\text{Re}\left(#1\right)}
\newcommand{\neff}{n}
\newcommand{\quoting}[1]{``#1''}

\begin{document}

\title{Goos-H\"anchen shift and localization of optical modes in deformed microcavities}
  \author{Julia Unterhinninghofen}
  \author{Jan Wiersig}
  \affiliation{Institut f{\"u}r Theoretische Physik, Universit{\"a}t Bremen,
  Postfach 330 440, D-28334 Bremen, Germany}
  \author{Martina Hentschel}
\affiliation{Max-Planck-Institut f\"ur Physik Komplexer Systeme, N{\"o}thnitzer
Stra{\ss }e 38, D-01187
Dresden, Germany}
\date{\today}

\begin{abstract}
Recently, an interesting phenomenon of spatial localization of optical modes 
along periodic ray trajectories near avoided resonance crossings has been
observed [J. Wiersig, Phys. Rev. Lett. {\bf 97}, 253901 (2006)]. 
For the case of a microdisk cavity with elliptical cross section we use
the Husimi function to analyse this localization in phase space. 
Moreover, we present a semiclassical explanation of this phenomenon in terms
of the Goos-H\"anchen shift which works very well even deep in the wave regime. This semiclassical correction to the ray dynamics modifies the
phase space structure such that modes can localize either on stable islands or
along unstable periodic ray trajectories.
\end{abstract}
\maketitle

\section{Introduction}
The phenomenon of avoided level crossings has been discovered more than 70
years ago by J. von Neumann and E.~P. Wigner~\cite{NW29}. Two energy
levels undergo an avoided crossing under variation of a control
parameter when they first approach very close to one another, and then depart
from each other without crossing.   
The energies $E_j$ of eigenstates in closed quantum-mechanical systems are
real-valued. In open systems, the energy eigenstates are replaced by
quasi-bound or resonant states with complex energies
$E_j$~\cite{Gamow28,KP38}. The imaginary part determines the lifetime $\tau_j
\propto 1/\imag{E_j}$ of the decaying  
state. The corresponding generalization of avoided level crossings are avoided
{\it resonance} crossings (ARCs)~\cite{Heiss00}. ARCs have been studied in a
number of physical settings, such as laser-assisted electron-atom
scattering~\cite{Kylstra98}, predissociation dynamics of
molecules~\cite{DLB95}, biased multiple quantum wells~\cite{WM93}, coupled
quantum dots~\cite{RS05}, quantum-dot microcavity
systems~\cite{Yoshi04,RSL04,PSMLHGB05}, microwave
cavities~\cite{PRSB00,DGH01}, and optical microcavities~\cite{WH06,Wiersig06}. 
In the latter systems the electromagnetic modes and their frequencies play the
role of eigenstates and their energies. 
One interesting phenomenon that can occur near ARCs is the
formation of fast and slowly decaying states~\cite{DJ95,PRSB00}.

In a recent letter it has been shown that this formation of fast and slowly
decaying states can be accompanied by a localization of the states along
unstable (or marginally unstable) periodic trajectories of the underlying classical
system~\cite{Wiersig06}. This localization is somewhat similar to the
well-known effect of scarring originally discovered for closed chaotic systems in the
field of quantum chaos~\cite{Heller84,Arranz98}.  

The aim of the present paper is twofold. First, we discuss the change of the
{\it phase space structure} of modes in optical microcavities when such
localization near an ARC occurs. The phase-space picture allows to clearly
identify the physical mechanism of lifetime enhancement (reduction): 
destructive (constructive) interference in real space leads to a reduction (enhancement) of intensity in the leaky region
of phase space. Second, we present a semiclassical explanation of the
appearance of localized modes near ARCs based on an augmented ray dynamics  
in which the Goos-H\"anchen shift (GHS) is included. The GHS is a lateral
shift of totally reflected beams along the optical interface~\cite{GH47}, i.e.~the points of incidence and reflection do not coincide. Rather, the beam appears to travel, as an evanescent wave, a short distance through the optically thinner medium. The GHS is proportional to the wavelength~$\lambda$. In the short-wavelength limit $\lambda\to 0$
the GHS therefore disappears leading to the standard ray dynamics of geometric optics.
In the field of quantum chaos, the short-wavelength limit corresponds to the
classical limit. For finite wavelength~$\lambda$ the GHS can be regarded as a
semiclassical correction to the ray dynamics. 
   
The paper is organized as follows. The system is defined in \sect~\ref{sec:system}. Section~\ref{sec:arc} presents the numerical analysis of localization of modes near ARCs. The semiclassical description in terms of the GHS is explained in \sect~\ref{sec:GHS}. Finally, a summary is given in ~\sect~\ref{sec:summary}. 

\section{The system}
\label{sec:system}
As model system we choose an optical microdisk with elliptical cross
section. Microdisks allow to confine light by total internal
reflection at the boundary which is highly relevant for various research
fields and applications~\cite{Vahala03}. Microdisks with deformed cross
section have attracted considerable attention in the context of quantum chaos
in open systems~\cite{ND97,GCNNSFSC98}. 
Elliptical microdisks have been studied both
theoretically~\cite{SRTCS04,SHI04} and experimentally~\cite{BHCK97,KKK04}. 
As semimajor axis we set $a = R(1+\varepsilon)$ and as semiminor axis $b =
R/(1+\varepsilon)$. The area of the ellipse $\pi ab$ does not depend on the
deformation parameter $\varepsilon$. The eccentricity is given by
$e=\sqrt{1-(b/a)^2} = \sqrt{1-1/(1+\varepsilon)^4}$. We choose  $R=1\mu$m and
the effective index of refraction $\neff = 3.3$ according to recent
experiments on circular microdisks~\cite{PSMLHGB05}. Maxwell's equations for
the transverse magnetic (TM) polarized modes $E_z(x,y,t) =
\psi(x,y)e^{-i\omega t}$ reduce to a two-dimensional  scalar wave
equation~\cite{Jackson83eng} 
\begin{equation}\label{eq:wave}
-\nabla^2\psi = \neff^2(x,y)\frac{\omega^2}{c^2}\psi \ ,
\end{equation}
with frequency $\omega$ and the speed of light in vacuum $c$. The wave
function $\psi$ and its normal derivative are continuous across the boundary
of the cavity. At infinity, outgoing wave conditions are imposed. With these
open boundary conditions the modes decay as light can leak out of the
cavity. This results in  quasi-bound states with complex frequencies
$\omega$ in the lower half-plane. The real part is the usual frequency. The
imaginary part is related to the lifetime $\tau=-1/[2\,\imag{\omega}]$ and to 
the quality factor $Q = -\real{\omega}/[2\imag{\omega}]$.

\section{Avoided resonance crossings}
\label{sec:arc}
\begin{figure}[ht]
\includegraphics[width=\figurewidth]{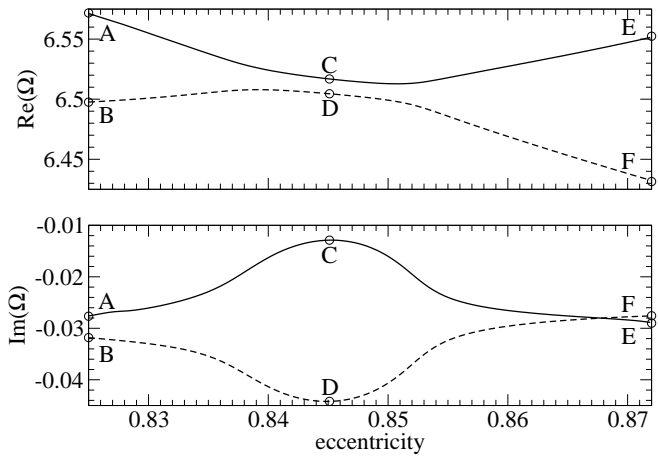}
\caption{An avoided resonance crossing in the elliptical microcavity. Plotted
 are the complex frequencies~$\Omega$ as function of the eccentricity. The
 real part shows an avoided crossing, whereas the imaginary part shows the
 formation of a long-lived mode~$C$ and a short-lived mode~$D$. Dots
 mark the frequency of the modes shown in Fig.~\ref{fig:modes_bowtie}.}
\label{fig:arc_bowtie}
\end{figure}

We compute the modes numerically using the boundary element 
method~\cite{Wiersig02b}. 
Figure~\ref{fig:arc_bowtie} shows the calculated normalized frequency $\Omega =
\omega R/c = kR$ as function of the  eccentricity~$e$. The considered region
around $\real{\Omega} = 6.5$ corresponds to a 
vacuum wavelength of about $970\,$nm. We observe that the real part of the
frequency performs an avoided crossing, and the imaginary part shows a
formation of a short-lived (large $|\imag{\Omega}|$) and a
long-lived state (small $|\imag{\Omega}|$). This is an ARC of external coupling
in the strong coupling regime~\cite{Wiersig06}. Observing an avoided crossing is the more remarkable when recalling that the corresponding
closed system with vanishing wave intensity on the boundary -- the
so-called elliptical billiard~\cite{WWD97,WWD98} -- belongs to the class of
integrable systems which typically do not show avoided level
crossings~\cite{Stoeckmann00}.  

\begin{figure}[ht]
\includegraphics[width=\figurewidth]{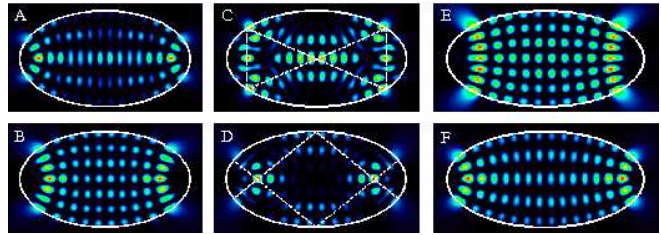}
\caption{(Color online) Calculated near field intensity of modes with the same
  labels as in Fig.~\ref{fig:arc_bowtie}. The long-lived \quoting{bowtie} mode $C$ and the short-lived mode $D$ show localization
  along periodic ray trajectories (dashed lines).}
\label{fig:modes_bowtie}
\end{figure}

The corresponding spatial mode structures are depicted in
Fig.~\ref{fig:modes_bowtie}. The modes $A$ and $B$ at the left hand side of
the ARC and the modes $E$ and $F$ at its right hand side look very
similar to the states in the elliptical
billiard~\cite{WWD97}. These states can be labelled
by two mode numbers $n_1$ and $n_2$. With $n_1$
($n_2$) we count the nodal lines in radial (azimuthal) direction in an
elliptical coordinate system~\cite{WWD97}. Note that, strictly speaking, the
wave function of a quasi-bound state does not possess nodal lines, only the real
and the imaginary part of the wave function does. For modes $A$ and $F$ we find
$(n_1, n_2) = (2, 26)$, and for modes $B$ and $E$ we find $(n_1, n_2) = (3,
18)$.

The hybridized modes $C$ and $D$, however, have no counterparts in the
elliptical billiard. They are superpositions of the original modes $A$ and $B$
(or $E$ and $F$) shown in Fig.~\ref{fig:modes_bowtie}. We can observe a clear 
accumulation of intensity on periodic ray trajectories. As these ray
trajectories are marginally stable~\cite{WWD97}, such kind of modes have been
referred to as \quoting{scarlike} in Ref.~\cite{Wiersig06}. True scars refer to  the localization of modes along unstable periodic ray trajectories in chaotic systems~\cite{Heller84}. It is customary to use the terminology "scars" also for systems with mixed phase space, see, e.g., Refs.~\cite{WFE96,SBF02,WRB05,HAD05,KKC06}.
Scarring has been observed also in microcavities~\cite{LLCMKA02,RTSCS02,FYC05,FHW06,WH08}. 

\begin{figure}[ht]
\includegraphics[width=\figurewidth]{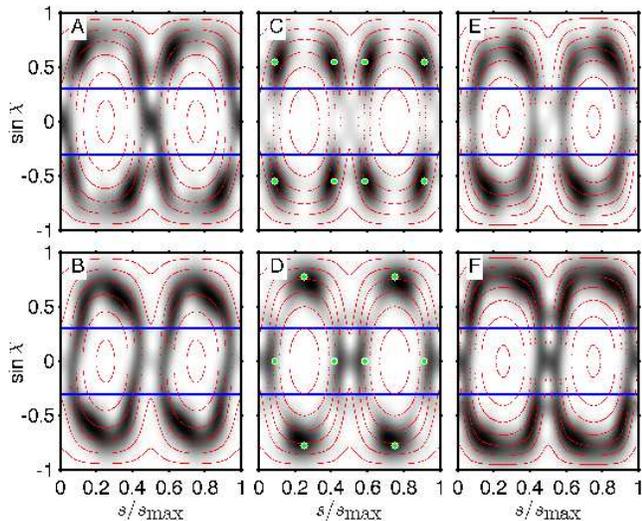}
\caption{(Color online) Emerging Husimi function (shaded regions) of the modes in 
Fig.~\ref{fig:modes_bowtie}. The horizontal lines are the critical lines
$\sin{\chi_c} = \pm 1/n$.  Thin dots mark some invariant curves in the phase
space of the ray dynamics. Thick dots represent the periodic ray trajectories 
on which the modes $C$ and $D$ are localized.} 
\label{fig:Husimi_bowtie}
\end{figure}

What causes the lifetime enhancement of the bowtie-shaped mode $C$ and the lifetime reduction of
mode $D$? In Ref.~\cite{Wiersig06} it was shown that the lifetime enhancement
(reduction)  of modes in rectangular cavities is due to destructive
(constructive) interference at dielectric corners. Scattering at dielectric
corners is the main decay channel in this kind
of cavities. For the elliptical cavity there are no corners. To understand
the lifetime modifications in this and other cavities without corners it is
convenient to analyse the (emerging) Husimi function~\cite{HSS03},
representing the wave 
analogue of the phase space. A phase space representation of 
the ray dynamics is the so-called Poincar\'e surface of section (SOS), as
shown in \fig~\ref{fig:Husimi_bowtie}. Whenever the trajectory leaves
the cavity's boundary after being reflected, its position~$s$ (arclength coordinate along the circumference) and tangential momentum~$p = \sin{\chi}$ (the angle of incidence $\chi$ is measured from the surface normal) is recorded. It is a
special property of integrable systems that each trajectory is confined to a so-called invariant curve (a torus in the full phase space). The invariant curve is exemplarily indicated by the thin dots in
Fig.~\ref{fig:Husimi_bowtie}. The symmetry properties of these curves in the SOS are determined by the discrete spatial symmetries of the billiard under investigation and the $p\to -p$ symmetry. The latter is due to time-reversal symmetry and hard-wall boundary conditions.   
The region between the two critical lines for total internal reflection
$\sin{\chi_c}=\pm 1/n$ is called the leaky region. A ray which enters this
region escapes according to Snell's and Fresnel's laws. 

From Fig.~\ref{fig:Husimi_bowtie} we can learn that far away from the ARC, the modes $A$, $B$, $E$, and $F$ follow roughly the ray dynamics as they are located near invariant curves. Note that the $p\to -p$ symmetry present in the corresponding billiard is broken in the open microcavity~\cite{LRKRK05,ADH08}. By comparing modes $A$ and $B$ with modes $C$ and $D$, it is apparent that near the ARC a rearrangement of the phase space structure takes place. The mode $D$ has an enhanced intensity inside the leaky region leading to a reduction of its lifetime. The mode $C$, however, has strongly suppressed intensity in the leaky region of phase space which increases its lifetime. Moreover, the phase space picture shows clearly the localization of the modes along periodic ray trajectories (thick dots in Fig.~\ref{fig:Husimi_bowtie}). 

\begin{figure}[ht]
\includegraphics[width=\figurewidth]{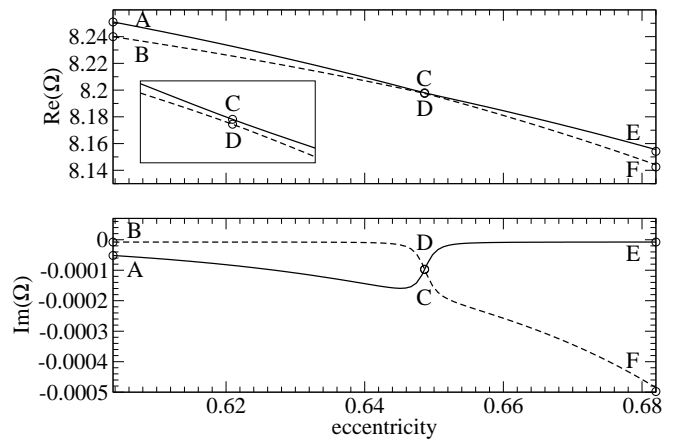}
\caption{An avoided resonance crossing with an avoided crossing in the real
 part of the frequency and a crossing in the imaginary part. Dots
 mark the frequency of the modes shown in Fig.~\ref{fig:modes_diamond}. Inset
 shows a magnification of the avoided crossing.}
\label{fig:arc_diamond}
\end{figure}

\begin{figure}[ht]
\includegraphics[width=\figurewidth]{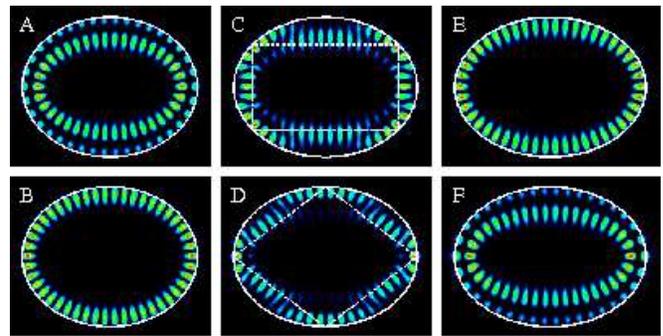}
\caption{(Color online) Calculated near field intensity of the modes with the
  same labels as in Fig.~\ref{fig:arc_diamond}. The rectangular-shaped mode $C$ and the diamond-shaped mode $D$ show
  localization along periodic ray trajectories (dashed lines).}
\label{fig:modes_diamond}
\end{figure}

In the following we demonstrate that the appearance of localized modes at ARCs
is not necessarily associated with the formation of short- and long-lived
modes. Figure~\ref{fig:arc_diamond} shows an example. Instead of the
formation of a long- and a short-lived mode we see a crossing in the
imaginary parts which is a signature of an ARC of internal type in the strong
coupling regime~\cite{WH06}.  
%
The corresponding mode patterns are depicted in Fig.~\ref{fig:modes_diamond}. 
The unperturbed modes $A$ and $B$ ($E$ and $F$) show a whispering-gallery 
structure similar to modes in a circular microdisk. The  mode numbers are 
$(n_1, n_2) = (1, 38)$ for mode $A$ and $(0, 46)$ for mode $B$. 
Also this type of ARC leads to hybridized modes, $C$ and $D$, having strong localization along periodic ray trajectories.  

\begin{figure}[ht]
\includegraphics[width=\figurewidth]{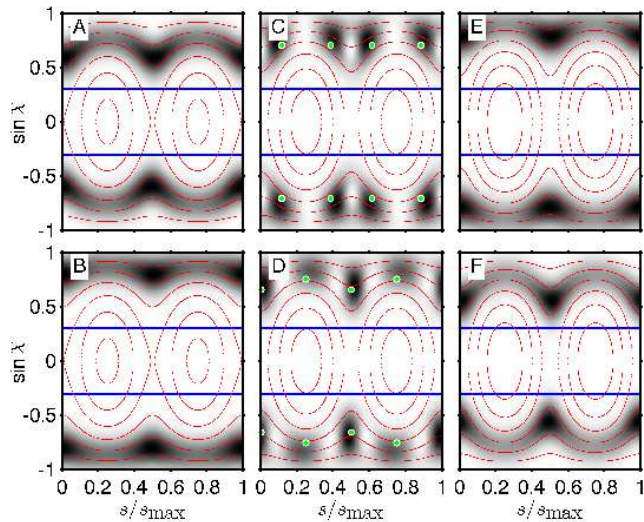}
\caption{(Color online) Emerging Husimi function (shaded regions) of modes in 
Fig.~\ref{fig:modes_diamond}. The horizontal lines mark the critical lines
$\sin{\chi_c} = \pm 1/n$.  Thin dots correspond to invariant curves in the
phase space of the ray dynamics. Thick dots represent the periodic
rays on which the modes $C$ and $D$ are localized.} 
\label{fig:Husimi_diamond}
\end{figure}

The localization is also clearly visible in the phase space representation in Fig.~\ref{fig:Husimi_diamond}. Moreover, this representation explains why there is no formation of short- and long-lived states in this case. Mode~$A$ lives closer to the leaky region than mode~$B$. Hence, the tail of mode~$A$ has a larger contribution inside the leaky region (even though the intensity is so weak that it cannot be seen in the figure). This explains why the lifetime of mode $A$ is much shorter (about a factor of 10) than the lifetime of mode~$B$. Due to their very different contributions in the leaky region a superposition of modes $A$ and $B$ cannot lead to a significant cancellation of phase space intensity in the leaky region. Consequently, short- and long-lived modes cannot form.  

\begin{figure}[ht]
\includegraphics[width=\figurewidth]{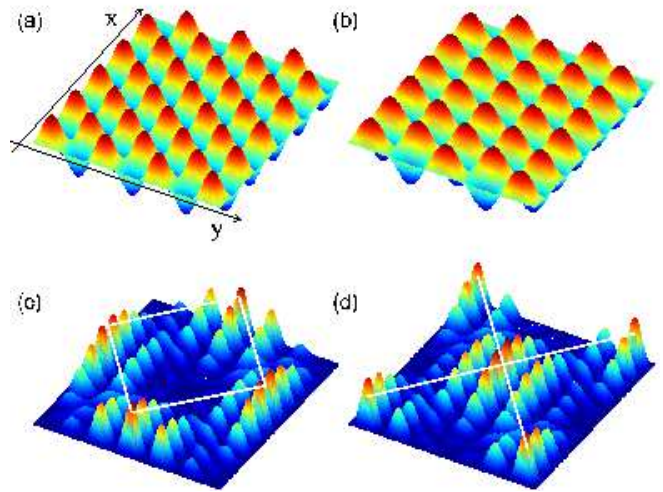}
\caption{(Color online) Eigenmodes and superpositions of eigenmodes in a rectangular
  billiard. (a) $\psi_{10,7}(x,y)$,  
  (b) $\psi_{12,5}(x,y)$, (c) $|\psi_{10,7}-\psi_{12,5}|^2$, and (d)
  $|\psi_{10,7}+\psi_{12,5}|^2$. Lines mark periodic ray trajectories.}
\label{fig:superpositions}
\end{figure}

The observed localization patterns can be explained for the cavities where the
corresponding billiard is integrable. The hybridised
modes, e.g. $C$ and $D$ in Fig.~\ref{fig:modes_diamond}, are superpositions of
the 
unperturbed modes far away from the ARC, e.g. $A$ and $B$
in Fig.~\ref{fig:modes_diamond}. As the eigenstates of an integrable billiard
the unperturbed modes $\psi$ have a simple structure of nodal lines. Therefore they can be labelled by a pair of mode numbers $(n_1,n_2)$ and $(m_1,m_2)$.
The morphology of superpositions $\alpha\psi_{n_1,n_2}+\beta\psi_{m_1,m_2}$ is
strongly influenced by the  
differences $|n_1-m_1|$ and $|n_2-m_2|$. If the differences are small numbers compared to $n_i$ and $m_i$,
then the mode pattern can be decomposed in a strongly oscillating part and a
weakly varying envelope. The weakly varying envelope defines the localization
pattern. Figure~\ref{fig:superpositions} illustrates this scenario for the
simpler case of a rectangular billiard. The modes can be computed analytically:
\begin{equation}\label{eq:billiard}
\psi_{n_x,n_y}(x,y) = \sin{\left(\frac{\pi
    n_x}{R}x\right)}\sin{\left(\frac{\pi n_y}{\varepsilon R}y\right)}
\end{equation}
if $0\leq x \leq R$ and $0\leq y \leq \varepsilon R$; otherwise
$\psi_{n_x,n_y}(x,y)=0$. The positive integers $n_x$, $n_y$ count the number
of nodal lines in $x$- and $y$-direction (boundaries $x=0$ and $y=0$ are not counted),
respectively. Figures~\ref{fig:superpositions}(a) and (b) show the cases
$(n_x,n_y) = (10,7)$ and $(m_x,m_y) = (12,5)$. It can be seen that the
superposition $\psi_{10,7}-\psi_{12,5}$ has reduced intensity at the
corner regions due to destructive interference. Correspondingly, the
intensity near the centre points is enhanced due to constructive
interference. The opposite is true for the superposition $\psi_{10,7}+\psi_{12,5}$. In both cases the localization resembles scarring on periodic
ray trajectories of short period. 


\section{Goos-H\"anchen shift}
\label{sec:GHS}
\begin{figure}[ht]
\includegraphics[width=0.8\figurewidth]{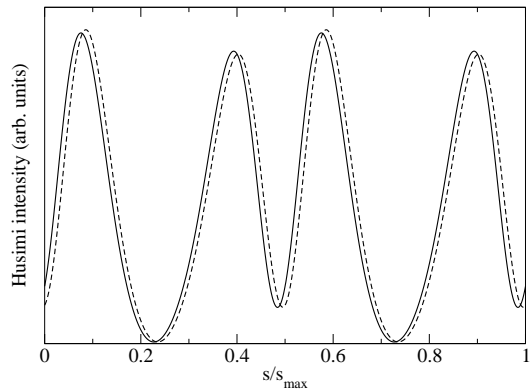}
\caption{Incident (solid line) and emerging (dashed) Husimi function of mode C in Fig.~\ref{fig:Husimi_diamond} for $p = \sin{\chi}= 0.75$.}
\label{fig:Inc_em_Husimi}
\end{figure}

In the previous section we demonstrated the existence of localized modes in optical microcavities near ARCs. But why is the light intensity in such a situation localized along periodic ray trajectories? To find an answer to this question we compare in Fig.~\ref{fig:Inc_em_Husimi} the incident and the emerging Husimi function~\cite{HSS03} for $p=0.75$. The incident (emerging) Husimi function represents the optical mode before (after) the reflection at the boundary of the microcavity. A lateral shift of about $\Delta s = 0.011 s_{\mbox{\footnotesize max}}$ can be observed. We interpret this shift as the Goos-H\"anchen shift (GHS). 

In the present section we develop a semiclassical description based on the GHS. We restrict ourselves to systems whose closed counterpart is integrable. Examples are the elliptical and the rectangular cavity. As well known in the field of quantum chaos, 
integrable closed systems typically show level crossings rather than avoided
level crossings. In the following we consider the crossing point of energy
levels where the levels form a  pair of degenerate levels. To do so, we use
the quantum mechanical notation for the solutions of the wave equation. 
We demonstrate that the degeneracy of energy levels implies degenerate
ray motion. Degenerate ray motion takes place on resonant tori which are
foliated by periodic ray trajectories, in contrast to generic tori which
support quasi-periodic motion. 

Let us consider two states each labelled by a pair of quantum numbers 
$(n_1,n_2)$ and $(m_1,m_2)$, respectively. In the limit of large quantum
numbers the semiclassical EBK quantization rule~\cite{Keller58} 
\begin{eqnarray}\label{eq:EBK}
E_{n_1,n_2} & = & H[(n_1+\frac{\alpha_1}{4})\hbar,(n_2+\frac{\alpha_2}{4})\hbar] \ ,\\
E_{m_1,m_2} & = & H[(m_1+\frac{\beta_1}{4})\hbar,(m_2+\frac{\beta_2}{4})\hbar]
\end{eqnarray}
is expected to give accurate results; $H$ is the classical Hamiltonian,
$\alpha_i$ and 
$\beta_i$ are Maslov indices~\cite{Maslov72}. Note that for systems with 
separatrices, Maslov indices can be defined only locally in phase space~\cite{RDWW96,WWD97,Wiersig01c}. However, if both states
belong to the same type of classical motion then $\alpha_i$ and 
$\beta_i$ are uniquely defined.   

Moreover, we assume in the following that $|n_i-m_i| \ll n_i, m_i$. In this
regime we can use the Taylor expansion in $\Delta n_i = m_i-n_i$
\begin{equation}\label{eq:taylor}
E_{m_1,m_2} \approx E_{n_1,n_2}+\frac{\partial H}{\partial I_1}\Delta n_1\hbar
+\frac{\partial H}{\partial I_2}\Delta n_2\hbar \ .
\end{equation}
In the case of degenerate levels $E_{m_1,m_2} = E_{n_1,n_2}$ we can write
Eq.~(\ref{eq:taylor}) with the frequencies $\omega_i = {\partial H}/{\partial
  I_i}$ as
\begin{equation}\label{eq:resonancecondition}
\omega_1\Delta n_1+\omega_2\Delta n_2 = 0 \ .
\end{equation}
Since $\Delta n_i$ are integer numbers, Eq.~(\ref{eq:resonancecondition})
requires that the frequency ratio $\omega_1/\omega_2$ is a rational number. As
a consequence the corresponding torus is resonant and composed of  
periodic ray trajectories. This line of arguments shows that the degeneracy of
quantum levels is related to the appearance of periodic ray trajectories for
the case of integrable closed systems.    

When an integrable system is subject to a smooth and generic perturbation 
then on the quantum (wave) side the degeneracy of the levels is lifted. As a
consequence, a level crossing turns into an avoided level crossing.
In the case of microcavities the perturbation is given by the open boundary
which allows for leakage of light out of the cavity.   
On the classical (ray) side, a generic perturbation breaks up the resonant
tori into a chain of stable periodic orbits (ray trajectories) surrounded by islands and unstable periodic orbits surrounded by a chaotic layer. This scenario 
is proven by the Poincar\'e-Birkhoff
theorem~\cite{Ott93,LichLieb92}. The fate of the nonresonant tori is determined by the
Kolmogorov-Arnol$'$d-Moser (KAM) theorem~\cite{Kolmog54,Arnold63,Moser62}. It
states basically that most of the nonresonant tori are conserved in the
perturbed system.
  
\begin{figure}[ht]
\includegraphics[width=\figurewidth]{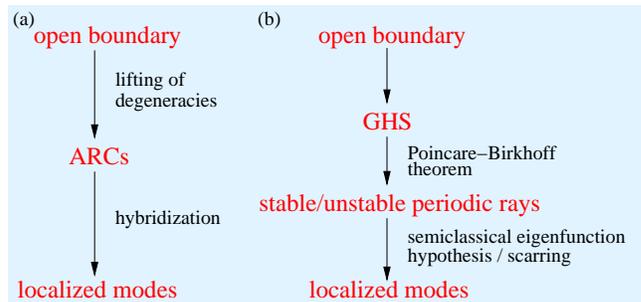}
\caption{(Color online) From the open boundary to localized modes near ARCs 
in systems with integrable closed counterpart. (a) Wave description and (b)
semiclassical description with augmented ray dynamics.} 
\label{fig:scheme}
\end{figure}

The basic idea here is to consider the GHS as a wavelength-dependent perturbation of the ray dynamics. The GHS is a direct consequence of the open boundary. The general scenario
is as follows. The GHS creates stable and unstable periodic orbits according to
the Poincar\'e-Birkhoff theorem. The size of the corresponding islands
surrounding the stable orbits in phase space depends on the ratio of
wavelength to cavity size. We get smaller islands for larger normalized frequencies $\Omega = kR$ and larger
islands for smaller $\Omega$. Following the semiclassical eigenfunction
hypothesis~\cite{Percival73,Berry77} the islands can support modes. In addition,
modes can be scarred along the unstable periodic ray trajectories. This semiclassical description is summarized in Fig.~\ref{fig:scheme} together with the wave description. It is important to note that the nonintegrability of the augmented ray dynamics explains the existence of ARCs.  

\begin{figure}[ht]
\includegraphics[width=\figurewidth]{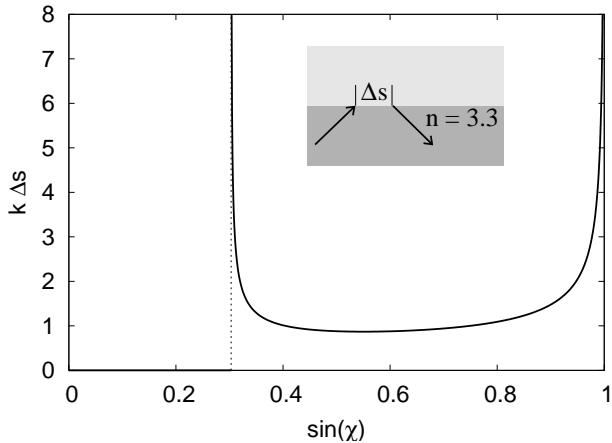}
\caption{Goos-H\"anchen shift $\Delta s$ at a planar interface according to
  Eq.~(\ref{eq:artmann}). The function $\Delta s$ is antisymmetric with respect
  to $\chi\to -\chi$, i.e. the shift measured along the direction of momentum is always non-negative. If $|\sin{\chi}| < 1/n$ then $\Delta s = 0$. The dashed
  line marks the critical line $\sin{\chi_c} = 1/n$.} 
\label{fig:artmann}
\end{figure}

The GHS $\Delta s$ for a TM polarized plane wave reflected at a planar dielectric interface is according to K. Artmann~\cite{Artmann51} given by 
\begin{equation}\label{eq:artmann}
k\Delta s = \realc{
\frac{2}{\sqrt{n^2\sin^2{\chi}-1}}\frac{\sin{\chi}}{\sqrt{1-\sin^2{\chi}}}} \ .
\end{equation}
The angle of reflection is equal to the angle of incidence $\chi$. Figure~\ref{fig:artmann} shows that the r.h.s. of Eq.~(\ref{eq:artmann}) has singularities at the critical angle $\sin{\chi_c} = 1/n$ and at $\sin{\chi} \to 1$. These singularities disappear in the case of incoming waves with finite spatial extension. For Gaussian beams of width~$\sigma$ Lai {\it et al.} derived an analytical expression~\cite{Lai86}. However, their result is restricted to $k\sigma \gg 1$, and exhibits artificial singularities in the low-$k\sigma$ regime considered here. Therefore, we use the Artmann result~(\ref{eq:artmann}) for long-lived ray trajectories which stay well within the interval $\sin{\chi} \in [0.35,0.9]$. Note that the GHS $\Delta s$ disappears in the semiclassical (short-wavelength) limit $k \to\infty$.
The GHS at curved interfaces has been computed numerically in Refs.~\cite{HS02,HS06} where, away from the critical angle in the regime of total internal reflection, reasonable agreement with the Artmann result has been found. No analytical formulas are available for the general case.

To compute the augmented ray dynamics we combine the billiard map
$(s_{j+1},p_{j+1}) = M(s_j,p_j)$ with the Goos-H\"anchen map $G(\lambda)$
defined by  
\begin{eqnarray}
s_{j+1} & = & s_j + \Delta s(p_j,\lambda) \ ,\\
p_{j+1} & = & p_j \ .
\end{eqnarray}
As the map $G(\lambda)$ preserves the phase space area, the combined map
$MG(\lambda)$ is Hamiltonian. In the short-wavelength limit we recover the
standard ray dynamics, i.e. $M = \lim_{\lambda\to 0} MG(\lambda)$.

\begin{figure}[ht]
\includegraphics[width=\figurewidth]{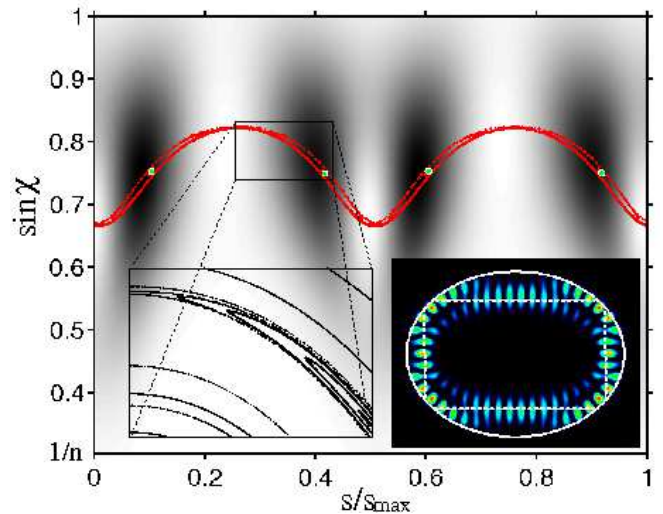}
\caption{(Color online) Emerging Husimi function of rectangular mode (mode $C$ in
  Figs.~\ref{fig:modes_diamond} and~\ref{fig:Husimi_diamond}) compared to
  the Poincar\'e SOS 
of the augmented ray dynamics above the critical line $\sin{\chi_c} = 1/n$. The thick dots mark the stable periodic ray
trajectory at the centre of the islands (thin dots). Left inset: magnification
  of one of the islands. Right inset: mode
structure and stable periodic ray trajectory in real space. Parameters are as 
in Figs.~\ref{fig:modes_diamond} and~\ref{fig:Husimi_diamond}.}  
\label{fig:GHS_diamond_C}
\end{figure}

Figure~\ref{fig:GHS_diamond_C} shows the phase space structure of the augmented ray dynamics for the situation of the rectangular mode at $e \approx 0.649$. The left inset shows that the GHS creates small islands in phase space. The center of these islands correspond to stable periodic orbits. The phase space in Fig.~\ref{fig:GHS_diamond_C} has been computed with Eq.~(\ref{eq:artmann}) for $\Omega = kR = 8.2$. It is important to mention that the overall island structure does not depend on $\Omega$; only the size of the islands does. The same island structure is also obtained when applying the formula of Lai {\it et al.} for Gaussian beams~\cite{Lai86} in the regime of large $k\sigma$ where the formula is valid.
From Fig.~\ref{fig:GHS_diamond_C} it can be clearly seen that the Husimi distribution is correlated with the island chain. Note that for both the Husimi distribution and the ray dynamics the outgoing component is shown, i.e. we plot the emerging Husimi distribution~\cite{HSS03} and the orbit just after reflection.
Figure~\ref{fig:GHS_diamond_C} reveals that the rectangular mode is not a scar with respect to the augmented ray dynamics but a mode living in a stable island chain following the semiclassical eigenfunction hypothesis. 
While the center of the Husimi function and the corresponding island agree well, the orientation of the two objects is different. This can be understand by the way the Husimi distribution is constructed. The Husimi distribution can be defined as Wigner distribution (restricted to the boundary of the cavity) folded with Gaussian wave packets of minimal uncertainty (coherent states)~\cite{HSS03}. The relative uncertainty and the orientation of the wave packets can be chosen freely, but usually the orientation is chosen in accordance with the $s$- and $p$-axes of phase space, as it is done in Fig.~\ref{fig:GHS_diamond_C}. However, one can also choose the orientation in accordance with the orientation of the island under consideration. Such "tilted coherent states" have been used, e.g., in~\cite{BKM05}. Moreover, note that the Husimi function for billiards and cavities itself is a semiclassical approximation~\cite{HSS03}.

The discussed ambiguity of the Husimi distribution is not present for the corresponding wave function in real space shown in the right inset of Fig.~\ref{fig:GHS_diamond_C}. In this real space representation the GHS is directly visible. The real space plot of the mode and the stable periodic ray trajectory shows much better agreement than in the case of the standard ray dynamics;
cf. Fig.~\ref{fig:modes_diamond}. In fact the agreement is surprisingly good
having in mind that we use the GHS for a planar interface and that $\Omega = 8.2$ is deep in the wave regime. 

The strong leaking of the Husimi function out of the islands in Fig.~\ref{fig:GHS_diamond_C} can be confirmed by computing semiclassically the number of modes $m$ fitting into an island of size $A$. From EBK quantization for the action $I = \frac{1}{2\pi}\oint p\,ds = m\hbar$ with $p = n\hbar k\sin{\chi}$ one gets 
\begin{equation}\label{eq:numberofmodes}
m = \frac{n\Omega s_{\mbox{\footnotesize max}}}{2\pi R} A(\Omega) \ ,
\end{equation}
where $A(\Omega)$ is the dimensionless phase space area spanned by $\sin{\chi}$ and 
$s/s_{\mbox{\footnotesize max}}$; the frequency $\Omega = kR$ is considered as real-valued for notational convenience. Numerically we find $A = 1.53\times 10^{-3}$ and $m = 0.04$. The value of $m$ appears to be too small, however, it is well-known that also small islands can accommodate modes. This is usually explained by the confinement due to cantori in the phase space region around the islands~\cite{WRB05,SLK08}. In our case the confinement is provided by the KAM tori (the unbroken tori of the elliptical billiard) surrounding the islands visible in the left inset of Fig.~\ref{fig:GHS_diamond_C}. 

\begin{figure}[ht]
\includegraphics[width=\figurewidth]{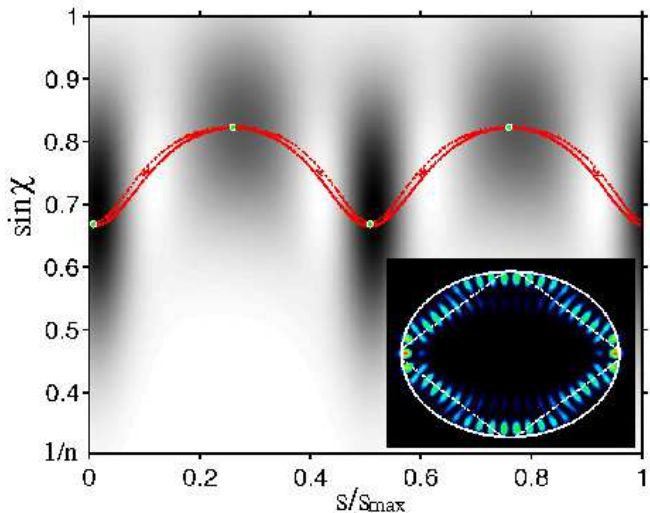}
\caption{(Color online) Emerging Husimi function of diamond mode (mode $D$ in
  Figs.~\ref{fig:modes_diamond} and~\ref{fig:Husimi_diamond}) compared to the
  Poincar\'e SOS 
of the augmented ray dynamics. The thick dots correspond the unstable periodic ray
trajectory. Inset: mode structure and unstable periodic ray trajectory in real
space. Parameters are as in Figs.~\ref{fig:modes_diamond}
and~\ref{fig:Husimi_diamond}.}  
\label{fig:GHS_diamond_D}
\end{figure}

Figure~\ref{fig:GHS_diamond_D} shows the same phase space as before but
together with the Husimi function of the diamond mode. Interestingly, this 
mode is localized on an unstable periodic ray trajectory of the augmented ray 
dynamics. This periodic ray trajectory is surrounded by a tiny chaotic layer which is not visible in Fig.~\ref{fig:GHS_diamond_D}. Hence, the diamond mode is
a true scar in the augmented ray dynamics. Again, the real space plot in the inset shows much better agreement than in the case of the standard ray dynamics;
cf. Fig.~\ref{fig:modes_diamond}

\begin{figure}[ht]
\includegraphics[width=\figurewidth]{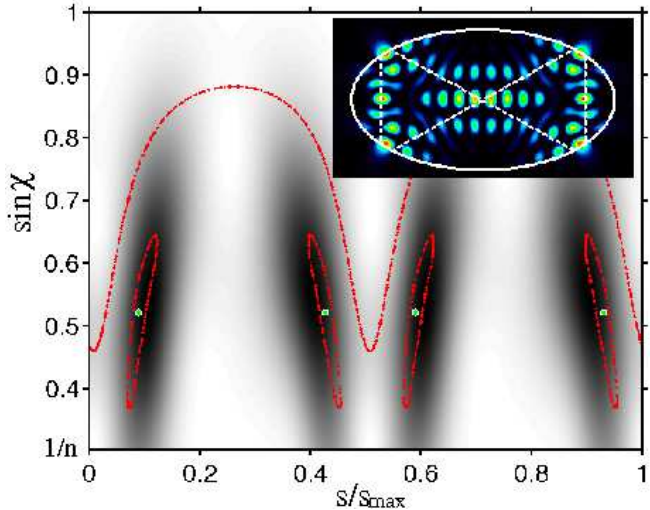}
\caption{(Color online) Emerging Husimi function of bowtie mode (mode $C$ in
  Figs.~\ref{fig:modes_bowtie} and~\ref{fig:Husimi_bowtie}) compared to the
  Poincar\'e SOS of the augmented ray dynamics. The
thick dots marks the stable periodic ray 
trajectory at the centre of the islands. Inset shows the mode
structure and the stable periodic ray trajectory in real space. Parameters are
as in 
Figs.~\ref{fig:modes_bowtie} and~\ref{fig:Husimi_bowtie}.}  
\label{fig:GHS_bowtie_C}
\end{figure}

Figure~\ref{fig:GHS_bowtie_C} eventually depicts the case of the bowtie mode $C$ with $e \approx 0.845$, 
cf. Figs.~\ref{fig:modes_bowtie} and~\ref{fig:Husimi_bowtie}. Again, the GHS
creates islands  in phase space. The inset of
Fig.~\ref{fig:GHS_bowtie_C} shows that the augmented ray 
dynamics fits slightly better the mode structure than the standard ray dynamics
shown in Fig.~\ref{fig:modes_bowtie}. 
In the case of Fig.~\ref{fig:GHS_bowtie_C} we find for the area $A = 4.7\times 10^{-3}$ and the number of modes in the islands $m = 0.11$.
The scenario for the hybridization partner, mode $D$, results in an unstable
periodic ray trajectory (not shown). Here, the GHS affects only the
overcritical reflections in the elongated region of the ellipse, whereas
the shift is zero at the remaining reflection points because of normal
incidence.

\begin{figure}[ht]
\includegraphics[width=\figurewidth]{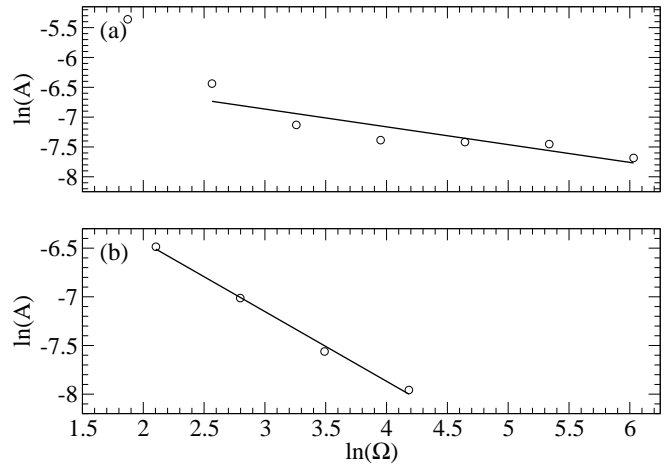}
\caption{Size of the island $A$ created by the GHS as function of $\Omega$ in
log-log plot; (a) eccentricity $e \approx 0.845$. The line is a linear fit
with slope $\approx -0.3$. (b) $e \approx 0.649$. The line is a linear fit with
slope $\approx -0.72$.}    
\label{fig:islandsize}
\end{figure}

Figure~\ref{fig:islandsize}(a) shows how the size of the islands scales with the frequency~$\Omega$ in the case of $e \approx 0.845$. For large enough $\Omega$ we see a power law dependence with exponent $\approx -0.3$. From Eq.~(\ref{eq:numberofmodes}) it
then follows that the number of GHS modes increase as $\Omega^{0.7}$ in the
semiclassical limit. Note that the first data point with $\Omega = 8.2$ is not
used in the fitting procedure since the corresponding island is strongly influenced by the critical line; see Fig.~\ref{fig:GHS_bowtie_C}.
For the case $e \approx 0.649$, Fig.~\ref{fig:islandsize}(b)
shows a power law with exponent $\approx -0.72$. Here the number of modes in
the semiclassical limit scales as $\Omega^{0.28}$. Interestingly, the exponent
depends on eccentricity. In both cases the number of GHS-induced modes increases as the semiclassical limit is approached (or the wavelength is fixed and
the cavity size is increased).

GHS-induced modes have been already predicted for the case  of 
microdome cavities~\cite{FCN07}. However, the relation to ARCs has not been
revealed in Ref.~\cite{FCN07}. 

\section{Summary}
\label{sec:summary}
We have analysed the phase space structure of localized modes formed near
avoided resonance crossings. Using the Husimi function we are able to
demonstrate the mechanism for the reduction (enhancement) of lifetimes due
to constructive (destructive) interference in real space leading to a reduction (enhancement) of intensity in the leaky region of phase space. 
We have developed a semiclassical extension of the ray model to describe the
wave phenomenon of localized modes for systems with integrable closed
counterpart. This augmented ray dynamics includes the Goos-H\"anchen
shift. As the Goos-H\"anchen shift breaks the integrability it naturally
explains both the appearance of avoided resonance crossings and the
localization of modes. According to the Poincar\'e-Birkhoff theorem the
perturbation creates stable and unstable periodic ray trajectories along which the modes localize. This scenario shows that half of the \quoting{scarlike} modes of Ref.~\cite{Wiersig06} can be regarded as true scars in the augmented ray dynamics. 
As example we discussed an optical microcavity with an elliptical cross
section, but we expect that our results apply to other cavity geometries as
well.
Moreover, we believe that the concept of an augmented ray dynamics can also be
applied to quantum systems, in particular to soft-wall
billiards~\cite{WRB05,KFA01}, in order to semiclassically describe quantum 
effects in a conceptually simple way.

\section{Acknowledgement}
We would like to thank A.~B\"acker, E.~Bogomolny, S.~W.~Kim, and H.~Waalkens
for discussions.  
Financial support by the DFG research group ``Scattering Systems with Complex
Dynamics'' and DFG Emmy Noether Programme is acknowledged.


\end{document}